\documentclass[final,1p,times,twocolumn,authoryear]{elsarticle}

\usepackage{amssymb}

\usepackage{lineno}

\usepackage{siunitx}
\usepackage{subcaption}

\journal{Nuclear Physics B}

\begin{document}

\begin{frontmatter}

\title{Prototype Cherenkov Detector Characterization for Muon Tomography Applications}

\author[affil]{T. Avgitas}
\ead{avgitas@ip2i.in2p3.fr}

\author[affil]{J.-C. Ianigro}
\ead{ianigro@ip2i.in2p3.fr}

\author[affil]{J. Marteau}
\ead{marteau@ip2i.in2p3.fr}

\affiliation[affil]{organization={IP2I, Univ Lyon, Univ Claude Bernard Lyon 1, CNRS/IN2P3, IP2I Lyon, UMR~5822},
            addressline={},
            postcode={F-69622},
            city={Villeurbanne},
            country={France}}

\begin{abstract}
Muography is an innovative imaging technique using naturally produced elementary particles -- atmospheric muons -- like the X-rays of medical imaging. The modification of the particles flux -- by scattering or absorption --, reflects the contrasts in density within the medium and therefore offers the possibility for an image of the crossed volumes. The imaging process is based on the tracking of the particles which accounts for the absorption or the scattering of the muons trajectories. Neither the energy nor the identity of the particles (the so-called PID) is exploited since this information traditionally relies on the use of calorimeters and/or high intensity magnetic fields. Both these techniques hinder detector portability which in the case of muography is important and this renders them impractical for its purpose. In this paper we characterize the performance of a simple and small water Cherenkov detector capable on the one hand of providing some insights on energy and PID and on the other hand of improving the background rejection for a muon telescope. We tested a prototype of such water Cherenkov detector in combination with two small muon hodoscopes. Both systems are using the same opto-electronics chain -- optical fibers and pixellized photosensors -- and the same data acquisition (DAQ) readout system which ensures an easy integration and implementation within presently running systems. This article presents the test setup, the detector response to cosmic muons and its performance evaluation against a basic simulation of its geometry and detection principle.

\end{abstract}




\end{frontmatter}


\section{Introduction}
\label{introduction}
Muon imaging or muography has emerged as a powerful non-invasive method to complement standard tools in Earth Sciences and is nowadays applied to a growing number of fields such as industrial controls, homeland security, civil engineering. This technique relies on the detection of modifications - absorption or scattering - in the atmospheric muon flux when these particles cross a target. Atmospheric muons are secondary products of primary cosmic-rays, essentially protons and helium nuclei expelled by stars, interacting with nuclei encountered on the top of the atmosphere. The rather low interaction cross-section of muons with matter ensures that most of them reach the Earth’s ground level and that furthermore they may significantly penetrate large and dense structures. As suggested originally by Alvarez in 1970 for the Chephren pyramid, this property may be exploited to perform density contrasts analysis of the interior of the target like X-rays do in medical imaging. Details on the basic muography features may be found in \citep{marteau_muons_2012} and references therein. 
Essentially muography relies on the tracking capabilities of portable detectors that are positioned behind (or on either side) of the structure under study. Many detection technologies are used so far : scintillators \citep{lesparre2012}, emulsions \citep{tioukov2019first} , resistive plate chambers \citep{ambrosino2015joint}, micro-megas \citep{bouteille2016} etc. They have various performances in spatial and angular resolutions which are in turn used to assess the absorption or the scattering efficiencies and then, after the resolution of an inverse problem, the density distributions in the target. But the tracking efficiency is not sufficient in principle when one aims to solve an inverse problem in muography, which is by essence constraint by the physics of muon-matter interaction. For instance, inverting data in scattering muography involves the momentum $p$ of the particle which enters in the multiple Coulomb scattering formula : $$\theta_0=\frac{13.6}{\beta c p}z\sqrt{\frac{x}{X_0}}\left[ 1 + 0.0038\ln\left(\frac{x}{X_0}\right)\right].$$
Inferring the particle momentum or energy is therefore a way to constrain the inverse problem algorithm to select the best solution among all the possible ones.  Since the measurement of energy/momentum of high energy particles requires detection capabilities well beyond the real possibilities of field equipment powered through solar panels for example. Exploiting the Cerenkov effect has been suggested already in the past \citep{pena2021} on the model of the AUGER experiment \cite{ref:AERA_Main}. The idea is to use water-based Cerenkov that may also provide interesting background rejection capabilities. In the MUTE experiment for example, a large water cuve is placed downstream from 2 scintillator planes. The typical dimensions of the cuve are of the ordre of 1m. The Cerenkov induced photons are collected by a large photomutiplier tube (PMT) while the scintillator planes are readout with SiPMs. 
In this article we present first results obtained with a prototype of water Cerenkov detector where the light collection is performed by a bundle of  WLS optical fibers connected via a mechanical cookie to a multi-anode PMT. This design offers multiple benefits among which the use of a compact readout system with no large size Photomultiplier Tube (PMT) such as the ones used in the AUGER experiment for instance. It allows also to re-use the same readout system as the one for the scintillators planes performing the tracking, with plastic scintillators planes of various transverse segmentation and active surface. This readout uses a custom, low power consumption, auto-triggerable Ethernet capable system described for instance in \citep{marteau_muons_2012}. Details are also given below. The paper describes the setup and the results obtained. It is organized as follows. The second and third sections describe the Cerenkov prototype and test setup. The fourth section details the track selection procedure while the fifth one gives the general outlines of the simulation tool developed for the data/MC comparison. Last section gives the main results and comment on the reasonable agreements achieved and opens on the perspectives for such detector design. 

\section{Cherenkov detector description}
\label{protCherDetDiscription}
\begin{figure}[h]
    \centering
    \includegraphics[width=0.80\textwidth]{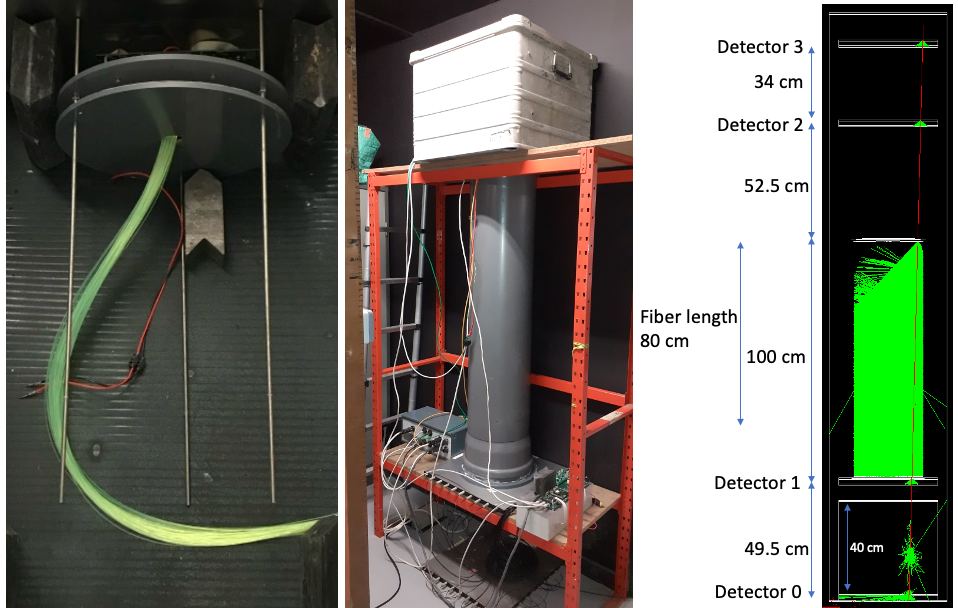}
    \caption{(left) The Cherenkov detector light collection system of fibers. The experimental setup (center) and its simulation counter part (right) for testing the DAQ and the Cherenkov response to atmospheric muons.}
    \label{fig:expSetup}
\end{figure}

For test purposes we designed the simplest possible tank for the Cherenkov detector prototype that consists of a cylinder of grey PVC with 1~cm thickness and 15.7~cm radius. The internal surface is covered with Tyvek to increase reflectivity, the water level inside stands at 1~m. A Hamamatsu H8804-300 Multianode Photomultiplier Tube (MaPMT) is used for the photon detection. The MaPMT window is optically coupled to 64 optic fibers (Kuraray-Y11) the length of which runs parallel to the cylinder's axis (fig. \ref{fig:expSetup}~-~left), their purpose is to increase the light collection efficiency by delivering the cherenkov photons to the MaPMT window.

The submerged part of the fibers is 80~cm in length, it has a rope-like structure with the fibers tangling with each other, something that should give rise to shadowing effects for the inner-most fibers of the bundle hindering collection efficiency and should be optimized in the future. This attribute is not a problem here since the purpose of this work was to test the feasibility of the DAQ, test how it operates in time coincidence with scintillation detectors and get insights on the capabilities, the potential drawbacks and guide future developments of the Cherenkov detector itself.

\subsection*{Characterization setup}
\label{characterSetup}
Two muon hodoscopes are placed on either side of the cherenkov detector, one top and one below (fig. \ref{fig:expSetup}~-~center/right). The top hodoscope is enclosed in a metal box to facilitate the alignment of the two detection planes. Each hodoscope consists of two detection planes that monitor the traversing particles' positions ($x$, $y$) with scintillator strips. The $z$ coordinate of the particle hits is provided by the planes' height positions since all four of them run parallel to the ground.

The top two planes consist of 16 scintillator strips per $x$ and $y$ direction with strip dimensions 2.5~cm~$\times$~40~cm and a distance of $\Delta z$~=~34~cm between planes while the bottom two planes are built with 8 scintillator strips per direction with 5~cm~$\times$~40~cm dimensions and set at a distance of $\Delta z$~=~49.5~cm between them. The DAQ setup and the readout electronics are similar to those used for the \textit{Diaphane} telescopes \citep{lesparre2012}, specifically for every detector, scintillator or Cherenkov, the photons are guided by WLS fibers to the window of a MaPMT. One MaPMT monitors the 64 scintillator strips for the top hodoscope and another one the 32 strips for the bottom hodoscope.

The \textbf{goal} of this work is to study the Cherenkov detector response to muon tracks crossing the scintillator surfaces, to develop a modelization of the detector and test in the future its potential energy and particle identification capabilities. To investigate this we performed two runs, one with lead bricks of 40 cm total height, between the bottom hodoscope planes and one without lead that we used as a reference. The \textbf{duration} of the first measurement (reference) was 453 hours and the second run, with the lead bricks, lasted 553 hours.

\subsection*{Triggering and track selection}
\label{muonTrackSelection}
The top hodoscope operates with an ``\textbf{and}'' trigger-0 condition between planes which means it triggers only when both planes register signal. The bottom hodoscope's trigger-0 condition is ``\textbf{or}'' which means that triggers when at least one out of the two planes registers signal. The trigger-0 condition for the Cherenkov detector is a simple voltage threshold applied by the electronics to the MaPMT signal. The trigger-1 condition that triggers the data registration for the entire experimental setup is set in ``\textbf{or}'' mode, which means that at least two trigger-0 conditions need to be satisfied. All the data discussed onward are only trigger-1 events. The reason behind this triggering schema is that we are interested in differentiating between particles that reach the rear bottom plane and those which stop in between planes for the bottom hodoscope (particle energy threshold detection).

The muons we are interested in are those that cross the entire volume of water inside the Cherenkov detector from the top water level to its bottom. We call these tracks ``\textbf{contained}'' to denote that they remain within the cylinder and they don't escape from the sides. The containment is determined from the interpolation of the reconstructed particle tracks to the height of the water level and the reconstructed track position on the detection plane 1 (fig \ref{fig:expSetup} - right) placed just below the cylinder.

\subsection*{Event selection}
\label{eventSelection}
A particle that crosses a detection plane should appear as a ``\textbf{hit}'' with ($x$, $y$) coordinates. This is not the case because at the level of the MaPMT there is crosstalk between channels and a particle crossing can trigger multiple channels. To mitigate this effect for the top MaPMT we apply a charge cut for all channels at the ``valley'' where the tail of the noise charge meets the rise of the photo-electron distribution. For the bottom MaPMT the DAQ threshold is high enough to cut these low charges. We treat the four detection planes of as unique hodoscope. The event selection process gives rise to three event categories:
\begin{itemize}
  \item \textbf{quads}, where all detection planes had at least one hit.
  \item \textbf{triples1}, where the three top planes had hits (Det. 1, 2, 3 - fig. \ref{fig:expSetup})
  \item \textbf{triples2}, where the top two planes and the rear bottom one had hits (Det. 0, 2, 3 - fig. \ref{fig:expSetup})
\end{itemize}

\subsection*{Track selection}
\label{trackSelection}
The track reconstruction for each event is done by performing a separate linear regression fit of the hits' projections on the $xz$ and the $yz$ planes. Events have different multiplicity of hits per hodoscope plane which lead to multiple track reconstructions for the same event. We define as a \textbf{track} the combination of hits that has the minimum chi-squared value for each projection plane. The chi-squared histograms for both directions ($x$ vs $z$ and $y$ vs $z$) are not all the products of tracks and they include also large values. We selected \textbf{chi-squared cut} value at the point where the distribution for the track populations change slope (in semi-log scale) which is around 7.81 for quads and at 5.99 for both triples populations. 

The reconstructed tracks sometimes happen to pass outside of the muon hodoscope borders. This in a small portion of cases is caused by inaccuracies in the exact positioning of the detectors which affect the reconstruction of the track. The most important effect though comes from the tracks of triples1 events. Extrapolating these tracks to the level of the last detector shows that the vast majority of these are actually tracks that are not detected because of the dimensions of the last detection plane. We mention this because there are also tracks that fall inside the last detection plane when extrapolated but are not detected due to the detection efficiency. We apply cuts for the reconstructed tracks position at the height of each detection plane from 0 to 40 cm in order to minimize this geometric effect.

The detection inefficiency is $\sim$10\% for each detection plane of the bottom hodoscope. This means that for the both measurements, with or without lead, triples2 events are due to the detection inefficiency of Det1 and for this reason we incorporated them into the quads sample. The same holds true for the triples1 measurements done with lead but it is impossible to separate between the part of the sample of triples1 that is due to the inefficiency of the Det0 detector and that caused by muons without enough energy to cross the lead bricks. We keep the triples1 tracks sample separate from the quads sample also for the measurements performed without lead to highlight the difference (see sec. \ref{results}).   
 
\section{Simulation}
\label{simulation}

\subsection*{Detector geometry}
\label{detGeom}
We developed a simplified simulation based on GEANT4 libraries (\cite{geant4_agost2003}, \cite{geant4_allis2006}, \cite{geant4_allis2016}) to evaluate the expected outcome for our characterization setup. The simulated muon flux assumes perfect alignment of the setup constituents around the $z$ axis, isotropy, and a complete shadowing of our detectors by the building floors on top. The first two assumptions have been already discussed to hold true only in approximation, the third one is true for the small zenithal angles corresponding to the contained tracks we use. The simulation is very simplistic but since the goal of this study was to test the DAQ setup and evaluate its performance for a Cherenkov detector we don't consider this a drawback.

\begin{figure}[h]
\centering
    \begin{subfigure}[b]{0.45\textwidth}
    \centering
    \includegraphics[width=\textwidth]{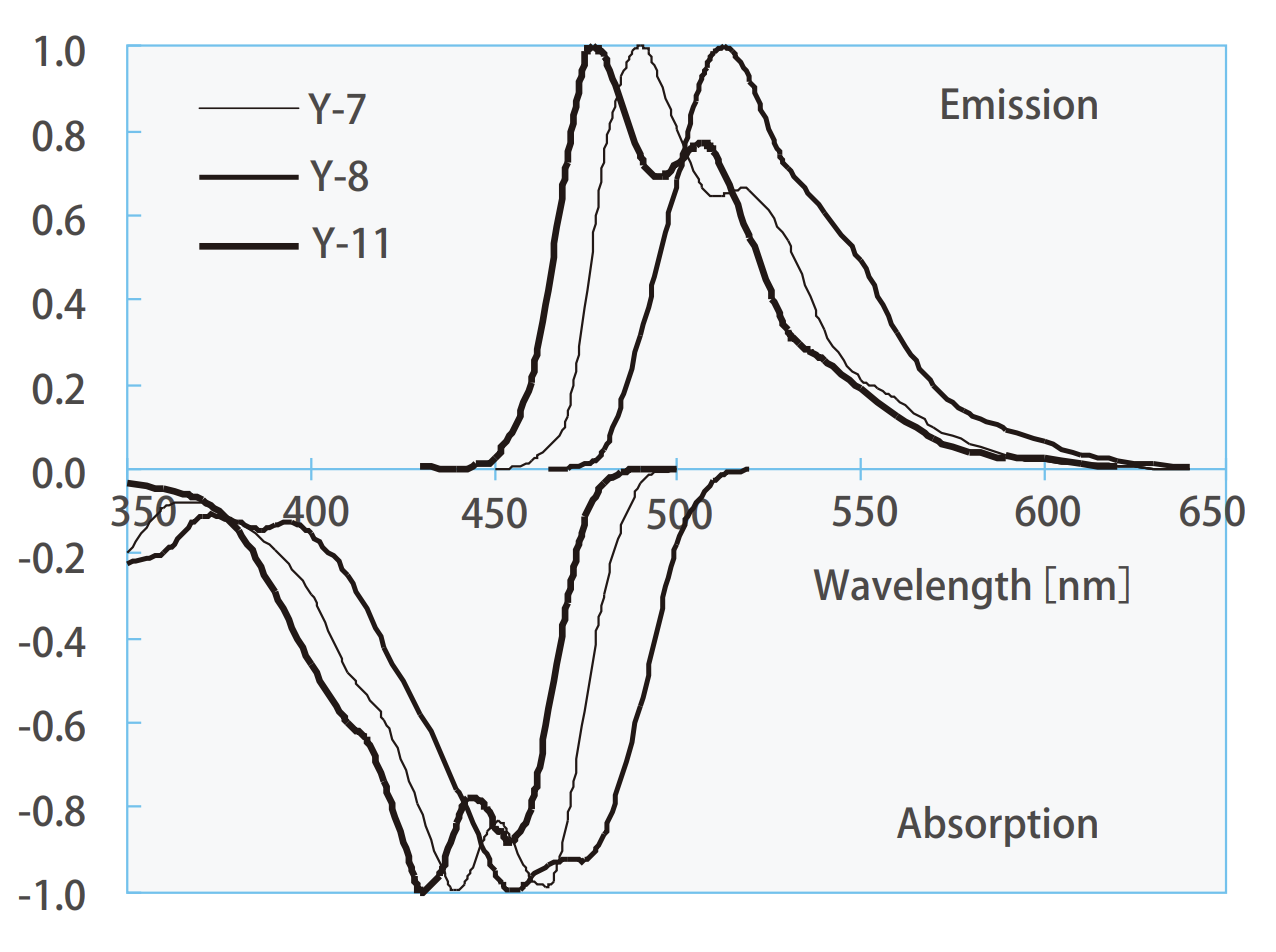}
    \caption{Y11 - WLS absorption spectrum}
    \label{fig:kurarayY11}
    \end{subfigure}
    \hfill
    \begin{subfigure}[b]{0.45\textwidth}
    \centering
    \includegraphics[width=\textwidth]{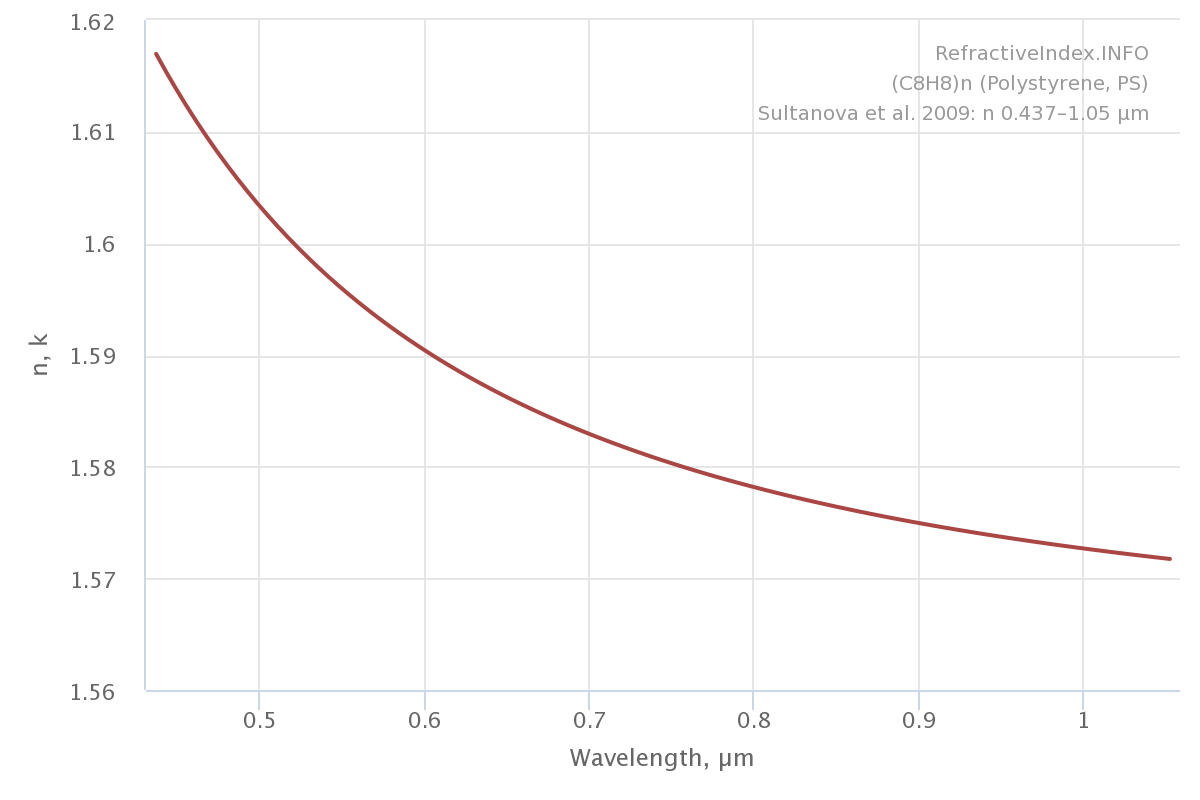}
    \caption{Polystyrene refractive index}
    \label{fig:polRefractiveIndex}
    \end{subfigure}
\caption{(left) Absorption and Emission Spectra for the WLS fiber used for the light collection of the Cherenkov detector (https://www.kuraray.com/uploads/5a717515df6f5/PR0150\_psf01.pdf). (right) polystyrene refractive index n (from https://refractiveindex.info/). }
\label{fig:absSpectraKurandRefractIndex}
\end{figure}

The \textbf{simulated} counterpart of the \textbf{Cherenkov detector} respects the geometric characteristics but fails to incorporate the optical properties of the materials. There is no reflection, refraction nor absorption for the photons traversing the water or coming in contact with the Tyvek surfaces. The same holds true for the optical fibers. The only thing taken into account is the wavelengths of the cherenkov photons and every photon that touches the optic fibers is considered detected provided that its wavelength falls within the limits of the absorption spectrum of the fibers (fig. \ref{fig:kurarayY11}). The distribution of the fibers is ideal, nothing like the true tangled geometry inside the actual detector, it respects the interaxial distance between fibers dictated by the mounting cookie throughout their entire length of 80~cm. The core of the optical fibers is made of polystyrene. We modeled the polystyrene refractive index dispersion ($n^2-1=\frac{1.4435 \lambda^2}{\lambda^2 - 0.02016}$, \cite{sultanova2009}, fig. \ref{fig:polRefractiveIndex}) so that we retrieve also the Cherenkov photon contribution from particles crossing the fibers themselves. This effect should be more pronounced for the actual entangled geometry of the fibers.

The \textbf{muon flux simulation} starts by producing Geantinos uniformly on the top plane of the bottom hodoscope with an isotropic azimuthal distribution. The zenithal angle is considered smaller than \ang{19} (based on the setup geometry) and the minimum energy to cross the building is 0.597~GeV. For the muon production we use the analytic formula found in \citep{shukla2018}. The Geantinos propagate upwards toward the roof of the building, they first need to cross a surface a little larger than that of detector 3 before finally reaching the area of the roof where they are ``killed'' and the information about the position, the direction and their energy is stored. The muon production incorporates this information to propagate them back towards the setup at the basement. For the analysis we use tracks that triggered both planes of the top hodoscope and at least one detection plane of the bottom hodoscope. For the simulation we use a trigger condition based on this and we include the Cherenkov detector response only for these events. 

\subsection*{Detector response}
\label{detResponse}
\begin{figure}[h]
    \centering
    \includegraphics[width=1.\textwidth]{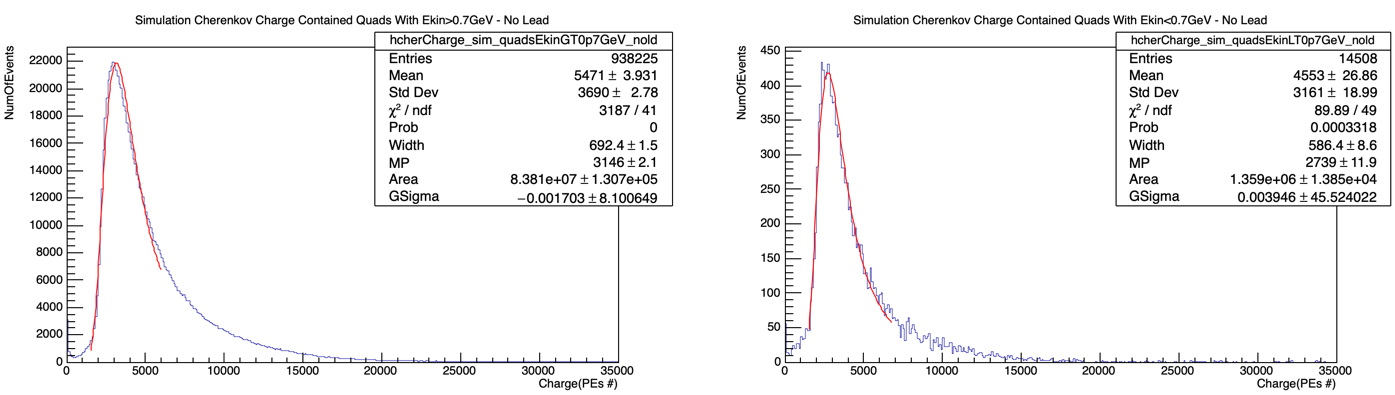}
    \caption{Cherenkov detector simulated response to two different muon energy intervals above (left)  and below (right) 0.7~GeV.}
    \label{fig:cherResponce4twoKinEnergies}
\end{figure}

The simulation shows that the Cherenkov detector's signal has a sudden peak followed by a very long tail. The peak can be described by a Gaussian and the largest proportion of the tail by an exponential but this needs to be studied further. Here we use a convolution of a Landau with a Gaussian distribution (a.k.a. Langau\footnote{https://root.cern.ch/root/html404/examples/langaus.C.html}) since this is the function we use also for the experimental data and we want to be consistent with our comparisons, in the future we will need to come up with a model for this. The first remark is that the 40~cm of lead have a small effect on the muon absorption, setting a minimum energy threshold for quads at $\sim$0.7~GeV, $\sim$500~MeV (from lead) + $\sim$200~MeV from water \citep{groom2001}. The expected detector response for these two energy intervals (below and above 0.7~GeV) is shown in fig.~\ref{fig:cherResponce4twoKinEnergies} in number of detected photo-electrons under the assumption of perfect efficiency.

We can safely assume from this that the differences for the Cherenkov response between contained quads and triples1 in the lead configuration will also come from differences in the muons energy and they will not be the product of just different angular distributions for the two populations. It also shows that there is a difference of 1000 photons for the mean values of the two histograms which becomes a more modest difference of 300 photons for the fit of the peaks. This may provide a weak insight on the energy of the particles and needs to be studied properly for the final detector geometry.

\section{Data/MC comparison}
\label{results}

\subsection*{Cherenkov detector response}
\label{cheDetResp}
In table~\ref{tab:detRespContainedEventsExp} we codify an overview of the results for the different measurements and the simulation. The key quantity we use is the most probable value position for the Langau. The percent column is the percentage between the number of events for a certain type and the total events detected. On the simulation level it gives some insight to the low energy particles effect on the Cherenkov detector response. On the experimental side the no-lead measurements show an inefficiency of the bottom hodoscope detector plane (Det1) at around 10\%. The triples2 for the experiment (already discussed) are the result of the Det0 efficiency and for both measurements sets are incorporated into the quads sample. The simulation shows clearly that the triples1 correspond to the low energy part of the muon spectrum. We see that the Langau peak is lower for these low energy muons and higher for high energy muons (quads) in both simulation and experiment.  

\begin{table}[h]
\centering
\begin{tabular}{|p{0.9cm}|p{0.9cm}|p{0.9cm}|p{2.3cm}|p{0.9cm}|p{0.9cm}|p{2.3cm}| }
 \hline
 \multicolumn{7}{|c|}{Detector Response for Contained Tracks - Experimental} \\
 \hline
 \multicolumn{1}{|c|}{} & \multicolumn{3}{|c|}{No Lead} & \multicolumn{3}{|c|}{Lead}\\
 \hline
 Type & Events & Percent & Peak Position & Events & Percent & Peak Position\\
 \hline
 quads    & 44976 & 88$\%$ & 29.95 $\pm$ 0.05 pe& 43405 & 71$\%$ & 29.57 $\pm$ 0.05 pe\\
 triples1 &  6290 & 12$\%$ & 29.06 $\pm$ 0.05 pe& 17929 & 29$\%$ & 26.62 $\pm$ 0.08 pe\\
 \hline
 total    & 51266 &  \multicolumn{2}{|c|}{} & 61334 & \multicolumn{2}{|c|}{} \\
 \hline
 \hline
 \multicolumn{7}{|c|}{Detector Response for Contained Tracks - Simulation} \\
 \hline
 \multicolumn{1}{|c|}{} & \multicolumn{3}{|c|}{No Lead} & \multicolumn{3}{|c|}{Lead}\\
 \hline
 Type & Events & Percent & Peak Position & Events & Percent & Peak Position\\
 \hline
 quads    & 951439 & 99.88$\%$ & 2939 $\pm$  5 pe & 929305  & 97$\%$ & 3050 $\pm$  5 pe\\
 triples1 &   1145 &  0.12$\%$ & 2710 $\pm$ 80 pe &  25746  &  3$\%$ & 2800 $\pm$ 22 pe\\
 \hline
 total    & 952584 &  \multicolumn{2}{|c|}{} & 955051 & \multicolumn{2}{|c|}{} \\
 \hline
\end{tabular}
\caption{Cherenkov detector response Langau peak position for different types of tracks - Simulation \& Experiment.}
\label{tab:detRespContainedEventsExp}
\end{table}

\subsection*{Angular distribution of tracks}
\label{angDistrTracks}
\begin{figure}[h]
    \centering
    \includegraphics[width=1.\textwidth]{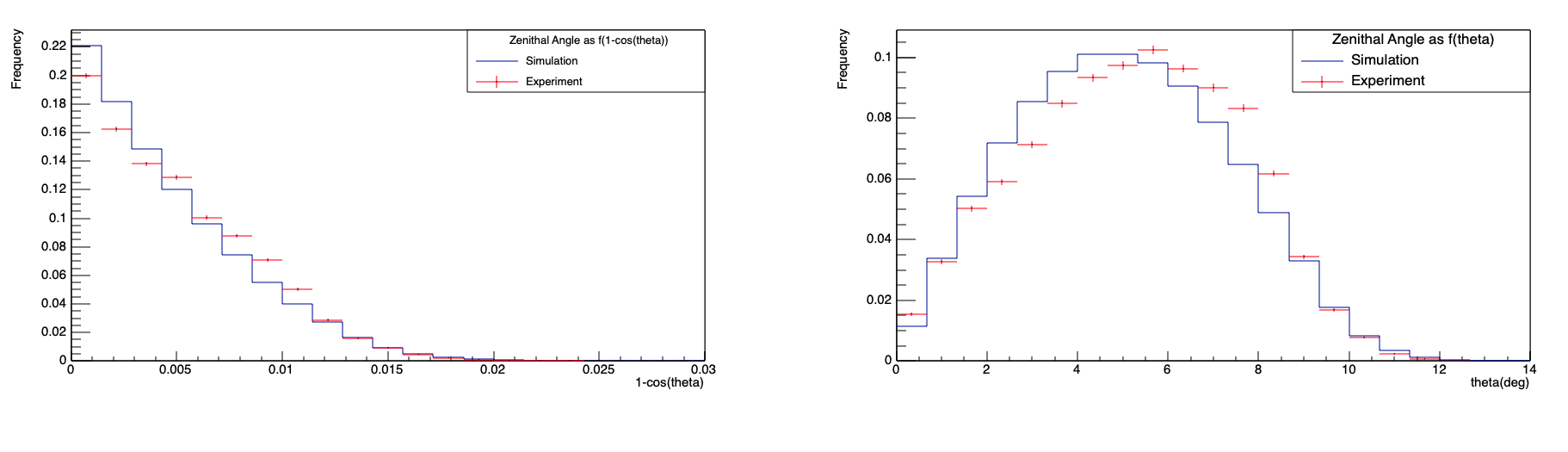}
    \caption{Contained Tracks - No Lead - Simulation Vs Experimental Data - Angular Distributions.\\
    Simulation (blue line) and experiment (red crosses) for the zenithal distribution (left) and the azimuthal distribution (right).}
    \label{fig:expvsSimNoLeadAngDistr}
\end{figure}

We provide the zenithal angular distribution ($\theta = 0\deg$ for vertical muons) for the muon tracks without lead in two representations (fig.~\ref{fig:expvsSimNoLeadAngDistr}). We see in both plots that the experimental data are skewed towards higher values, an effect that is more pronounced in the right plot. We attribute this divergence of the experimental data to the distribution of matter for the building above which is very difficult to implement properly for the Monte-Carlo simulation. The overall normalization though is correct.

\subsection*{Low energy events}
\label{lowE}
\begin{figure}[h]
\centering
    \includegraphics[width=1.\textwidth]{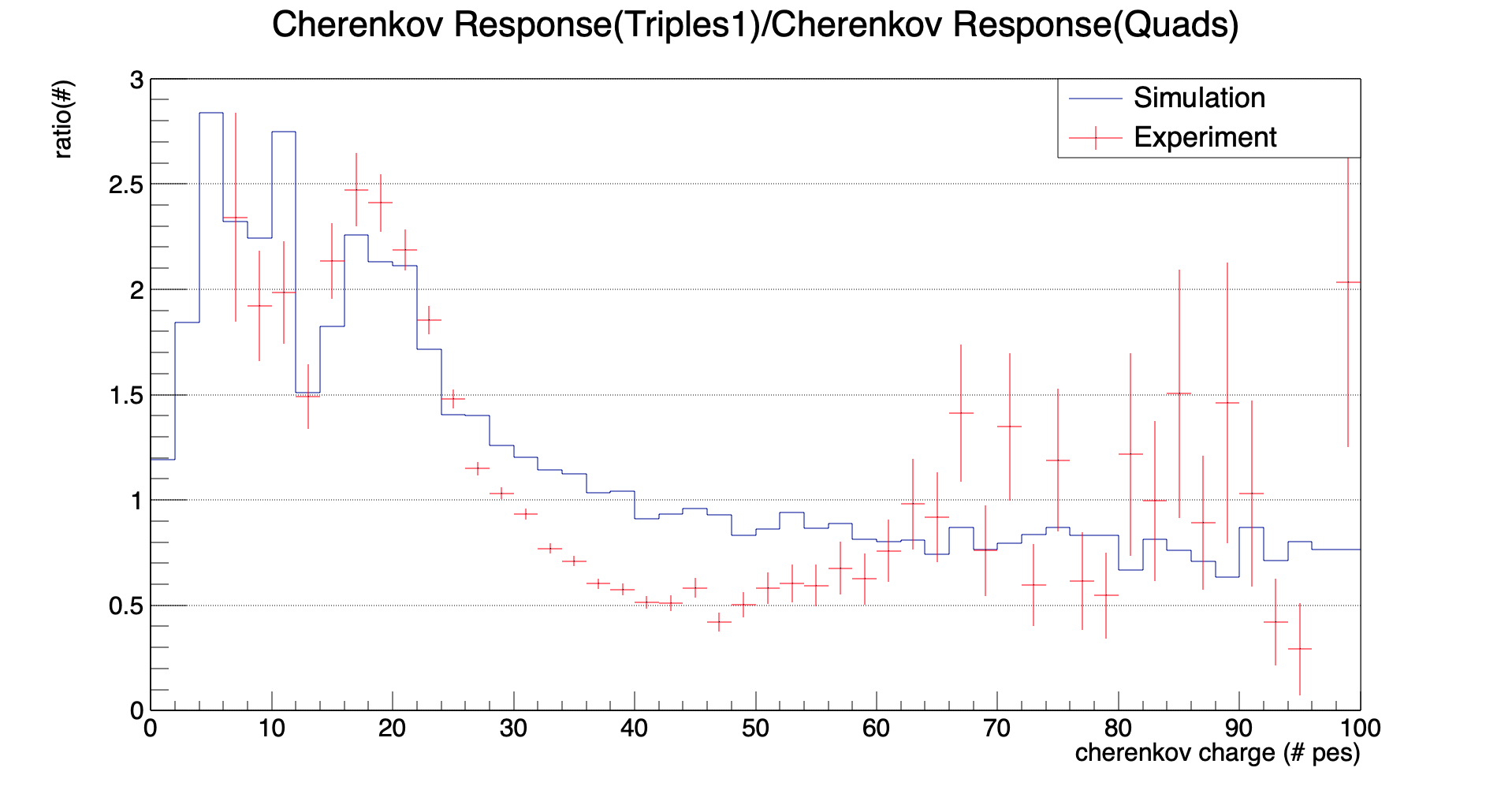}
    \caption{Contained Tracks - No Lead - Simulation Vs Experimental Data - Angular Distributions.\\
    Simulation (blue line) and experiment (red crosses) for the zenithal distribution (left) and the azimuthal distribution (right).}
    \label{fig:expvsSimRatios}
\end{figure}

In the experimental setup with lead between the bottom hodoscope's detectors the response of the cherenkov detector to quads (\& triples2) corresponds to muons with higher energies (\textgreater0.7~GeV) while the triples1 correspond to the lower part of the spectrum. The ratio of the cherenkov response to triples1 and quads shows (fig.~\ref{fig:expvsSimRatios}) that the detector differentiates for the low energy part of the spectrum and the high energy part of the spectrum at 26 - 30 pes, where the charge ratio becomes $\sim$1. The probability that the cherenkov response to low energy muons is lower than 26 pes is 1.5 to 2 times higher than for the high energy tracks to give signal at that region. This result is consistent with the simulation expectation. 

We see a similar behavior but reversed and less exaggerated for the charge response region above 55 pes. Here the probability (on a bin by bin basis) that the lower energy muons would give signal at this region is 20\% less than for the high energy ones. The behavior of the experimental results for the cherenkov response region is also consistent with the expectations from the Monte-Carlo.

Finally there is a region, between 30 pes and 50 pes, for which the agreement between simulation and experiment is only qualitatively similar. The simulation shows that there is a very small probability of differentiating between low energy and high energy events in contrast to the detector that seems to give up to 2 times more chance of the high energy particles to give charge in this region. This is something that we need to investigate further for the future iteration of the cherenkov detector design.

\section{Conclusions}
\label{conclusions}

This article presents the design of a portable, compact, robust and low power water-Cerenkov detector to be used in field muography experiments. The results obtained show that Cerenkov photons may be collected by a bundle of WLS fibers and amplified by the same MaPMT than those used for the scintillator readout. This simple particular design offers many possibilities for the fibers layout in the water tank and for the integration of the Cerenkov sub-detector in any kind of scintillator tracker. 
The use of the Cerenkov information gives access to inputs on the incident particle energy, as we demonstrated with a simple high-energy pass filter made of lead bricks in our setup. The present study allowed to validate a design and its Monte-Carlo model. These inputs may be of primary importance to better constrain the reconstruction algorithms of the muography imaging especially in scattering mode. 
Next steps are the production of an optimized detector where the layout of the fibers is upgraded to have a maximal Cerenkov light collection efficiency for downward propagating particles (ie the muon crossing the target) and a detailed study of the muon/electron separation power. The newly designed Cerenkov detector will be implemented on running muon stations where it will replace the passive lead shielding used to remove low-energy background at present. This implementation fits with the present mechanical frame and with the already running DAQ system.

\section*{Acknowledgements}
This work has been carried out with the support of the LabEx LIO of the Université Claude Bernard Lyon 1, which has been created as part of a program for future development (reference number ANR-10-LABX-066) initiated by the French government and overseen by the Agence Nationale de la Recherche (ANR). This study is part of the ANR-19-CE05-0033 MEGAMU project. 





\end{document}